\documentclass[5p,times,twocolumn]{elsarticle}

 \usepackage{graphics}
\usepackage{graphicx}
 \usepackage{epsfig}

\usepackage{amssymb}
\usepackage{amsthm}

 \usepackage{lineno}

\biboptions{comma,square}

\begin{document}

\begin{frontmatter}






\title{The X-Array and SATURN: A new decay-spectroscopy station for CARIBU}

\address[UML]{Department of Physics and Applied Physics, University of Massachusetts Lowell, Lowell, MA 01854, USA}
\address[ANL]{Physics Division, Argonne National Laboratory, Argonne, Il 60439, USA}
\address[NED]{Nuclear Engineering Division, Argonne National Laboratory, Argonne, Il 60439, USA}
\address[UOC]{Department of Physics, University of Chicago, Chicago, Il 60637, USA}
\cortext[cor1]{Corresponding Author: A.J.~Mitchell, Department of Physics and Applied Physics, University of Massachusetts Lowell, Lowell, MA 01854, USA; Email, Alan$\_$Mitchell@uml.edu; Phone, +1 978 934 3989}
 \fntext[NASA]{Present address: Marshall Space Flight Center, Huntsville, AL 35812}
 \fntext[IIT]{Present address: Department of Physics, Indian Institute of Technology Roorkee, Roorkee 247 667, INDIA}
 \fntext[BNL]{Present address: National Nuclear Data Centre, Brookhaven National Laboratory, Upton, NY 11973, USA}
\author[UML]{A.J.~Mitchell\corref{cor1}}
\author[ANL]{P.F.~Bertone\fnref{NASA}}
\author[ANL]{B.~DiGiovine}
\author[UML]{C.J.~Lister}
\author[ANL]{M.~P.~Carpenter}
\author[UML]{P.~Chowdhury}
\author[ANL]{J.A.~Clark}
\author[UML]{N.~D'Olympia}
\author[UML]{A.Y.~Deo\fnref{IIT}}
\author[ANL,NED]{F.G.~Kondev}
\author[ANL]{E.A.~McCutchan\fnref{BNL}}
\author[ANL]{J.~Rohrer}
\author[ANL,UOC]{G.~Savard}
\author[ANL]{D.~Seweryniak}
\author[ANL]{S.~Zhu}



\begin{abstract}

A new decay-spectroscopy station has been commissioned for experiments with low-energy, fission-fragment radioactive beams from the CARIBU ion source. The new set-up consists of the `X-array', a highly-efficient array of HPGe clover detectors, and `SATURN' (Scintillator And Tape Using Radioactive Nuclei), a plastic scintillator detector combined with a tape-transport system for detection of $\beta$ particles and removal of long-lived isobaric decay products. \\

\end{abstract}


\end{frontmatter}





\section{Introduction}\label{introduction}

Exploration of the nuclear landscape far from stability is contingent on major advancements in generating isotopically pure beams of exotic nuclei. The  CAlifornium Rare Isotope Beam Upgrade (CARIBU \cite{Savard:2008}) to ATLAS at Argonne National Laboratory (ANL) is an example of such a facility. CARIBU utilizes the spontaneous fission branch of $^{252}$Cf to produce exotic neutron-rich beams. The availability of such beams allows for experiments in regions of the nuclear chart that have until now been difficult to access, thereby addressing a broad variety of topics within nuclear structure, nuclear astrophysics and nuclear physics applications. An extensive research program utilizing CARIBU low-energy beams is about to commence. For many experiments the nuclides of greatest interest will be produced with very low cross sections, resulting in beam intensities lower than hundreds of ions per second, therefore building efficient experiments is of paramount importance. A new decay spectroscopy station, optimized for performing very efficient $\beta-\gamma$ coincidence measurements with low-energy CARIBU beams, has been designed, installed and commissioned. The set-up includes a highly-efficient array of High-Purity Germanium (HPGe) `clover' detectors in conjunction with plastic scintillators for $\beta$-particle detection. To avoid accumulation of activity from the decay chain following the beam species of interest, a moveable tape has been incorporated into the system. This arrangement allows for beam collection directly on the tape at the geometrical centre of the detectors. Implanted nuclei are allowed to decay for a fixed period of time, before removal to a point well shielded from the detectors. The authors hope that this article will serve as a useful tool to those who wish to utilize the new decay-spectroscopy station in the forthcoming experimental campaign and beyond.  


\section{Design concepts}\label{design}


\subsection{$\gamma$-ray detection}\label{detection}

In decay studies of very exotic nuclei, in which beam intensities and $\gamma$-ray multiplicities are low, achieving the highest possible detection efficiency is vital. The optimum experimental arrangement is one by which large, HPGe detectors with high intrinsic efficiency offer extensive solid angle coverage, thus ensuring the maximum probability of detecting the few $\gamma$ rays that are emitted. An effective manner of doing so is to place the detectors in a simple `box-like' geometry close to the beam collection point, thus ensuing large solid-angle coverage. A HPGe-detector array with these characteristics, the `X-Array' \cite{Moore:2003}, has been developed and implemented at ANL for such experiments. Until recently, the X-Array has primarily been utilized in conjunction with a double-sided silicon strip detector (DSSD) located at the focal plane of the Fragment Mass Analyzer (FMA \cite{Davids1989}). However, flexibility in the design of the X-Array allows for its use in decay spectroscopy with low-energy CARIBU beams. 

There are several features that must be considered with very close geometrical arrangements of clover detectors. These include Compton scattering between crystals (within the same cryostat or between clovers), `summing' of two or more photons from the same decay hitting a single crystal, and the effects of high `chance' coincidence rates. Careful analysis of the data can ameliorate some of these problems and even improve the efficiency and spectral quality, and is discussed in Section \ref{performance}. As the $\gamma$-ray multiplicity associated with each $\beta$ decay is relatively low, $\sim$~1-2, and the number of individual crystals is high, 20, the benefits of close geometry outweigh potential drawbacks. For special experiments where summing is an issue, the clovers can be moved further from the beam collection point and passive shields inserted between the cryostats. 


\subsection{$\beta$-particle detection}\label{bdetection}

In addition to high-efficiency detection of $\gamma$ rays it is desirable to detect $\beta$ particles emitted in the decay process. This enables suppression of background environmental $\gamma$ radiation though application of coincidence conditions, provides an initial time-of-decay signal and can enhance channel selection through selection of specific $\beta$-particle energies. To this end, a series of plastic-scintillator detectors, designed specifically for use with the X-Array and low-energy CARIBU beams, have been commissioned. Signals from the $\beta$ counter provide an optional trigger for the data acquisition system. By applying timing conditions to the detected $\gamma$ rays, it is possible to generate $\beta$-gated $\gamma$ and $\gamma\gamma$ coincidence spectra.

Two different designs have been implemented. The `Mark-1' detector is a stand-alone aluminum vacuum chamber which is mountable to the CARIBU beamline. The detector volume is designed fit within the open space between the X-Array clovers, with the beam deposited on aluminum foil located at its geometric centre. Both a large-volume, cylindrical `well' detector and series of thin paddles of the same scintillating plastic have been manufactured. The thickness of the well detector is large enough so that $\beta$ particles up to $\sim 10$~MeV are fully stopped in the material. As the full energy of $\beta$ particles emitted in a wide range of decay processes is fully deposited in the detector, $\beta$ spectra and decay end-points can be determined by gating on individual $\gamma$ rays. A trade off in possessing this capability is a reduction in $\gamma$ efficiency of the X-Array, by approximately a factor of two. Also, with a single plastic counter, it is not possible to determine any directional information regarding the emission of $\beta$ particles with respect to their coincident $\gamma$ rays. The paddle detectors aim to circumvent these limitations. Each paddle is coupled to a separate photomultipler tube and so a degree of spatial information is obtained. Each paddle is positioned in front of a  Ge clover so $\gamma$ decays parallel, at 90$^{\circ}$ and antiparallel to the $\beta$ particle can be detected. With much thinner plastic, $\gamma$ attenuation is significantly reduced yet retains the ability to utilize $\beta$ signals as a trigger.


\subsection{Removal of long-lived radiation}\label{removal}

An issue associated with a simple `beam on'-`beam off' collect-and-measure approach is that over time, long-lived products of the subsequent decay chain accumulate and eventually dominate the data being collected. This particularly becomes an acute problem when studying nuclei far from stability, in which the decay chain is extensive and its members have significantly longer half-lives than the nucleus of interest. One approach to dealing with this problem is to collect the incoming beam on a movable tape and periodically remove the collected activity to a point which is well shielded from the detectors. Tape-transport systems of varying design have been utilized in nuclear physics research for many years with the general purpose of transporting radiation to or from a detector, depending on the requirements of the experiment. A tape-transport system for removing long-lived decay products has been incorporated into the design of the second $\beta$-detection chamber. SATURN (Scintillator And Tape Using Radioactive Nuclei) is largely based upon a Louisiana State University design \cite{Mlekodaj:1981} that makes use of a continuous loop of tape, and an initial test station that is located on one of the CARIBU low-energy beamlines. The primary focus of our research with this apparatus is intended to be with short-lived activities, such as in the $^{132}$Sn region, with half-lives of the order of a second or less. This has been factored into the design of SATURN so that it is capable of rapid tape movement within a short time period.


\section{Construction}\label{construction}


\subsection{The CARIBU low-energy beamline}\label{beamline}

\begin{figure}[h!]
\begin{center}
\includegraphics[width=8.5cm]{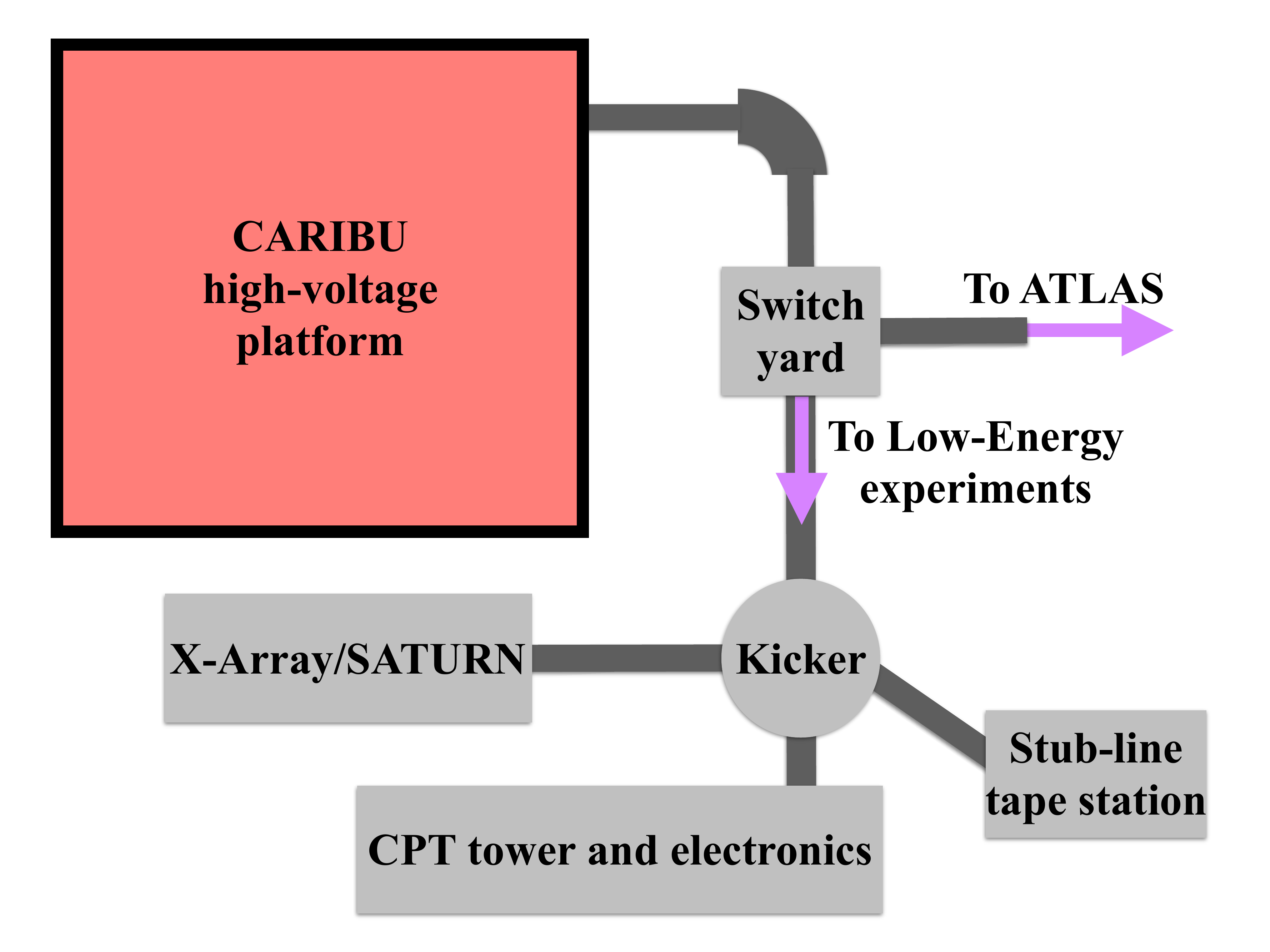}
\caption{\label{fig:CaribuHall} (Color online) Schematic diagram of the experimental equipment located in the CARIBU Low-Energy Hall.}
\end{center}
\end{figure}

CARIBU utilizes $^{252}$Cf, which has a 3.1$\%$ spontaneous fission branch and half-life of 2.64~years, in generating beams of neutron-rich radioactive nuclei. Fission fragments are extracted from the source and thermalized in the gas catcher via a combination of collisions within high-purity He gas and electromagnetic forces. The beam species of choice is extracted by the isobar separator with mass resolution $> 1:10,000$ and is directed towards a switch yard where it can either be delivered to ATLAS via a charge breeding process, or guided towards the low-energy experimental area. A new quadrupole triplet has recently been installed to improve transmission optics of the low-energy beamline. The beam is cooled and bunched in an RFQ buncher, from which ion pulses are released at 100-ms intervals. At this point it is also transformed from $\sim50$~keV to $\sim2$~keV by an `elevator'. The beam is deflected by a double electrostatic `kicker' to one of three possible experimental stations: the `stub-line' diagnostic tape station, the Canadian Penning Trap or the X-Array (see Fig.~\ref{fig:CaribuHall}). The kicker also functions as a beam sweeper when required. The deflectors are located $\sim 4$~m from the X-Array and so deflected beams do not contribute to the count rate. 

\subsection{The X-Array}\label{xarray}

\begin{figure}[h!]
\begin{center}
\includegraphics[width=8.5cm]{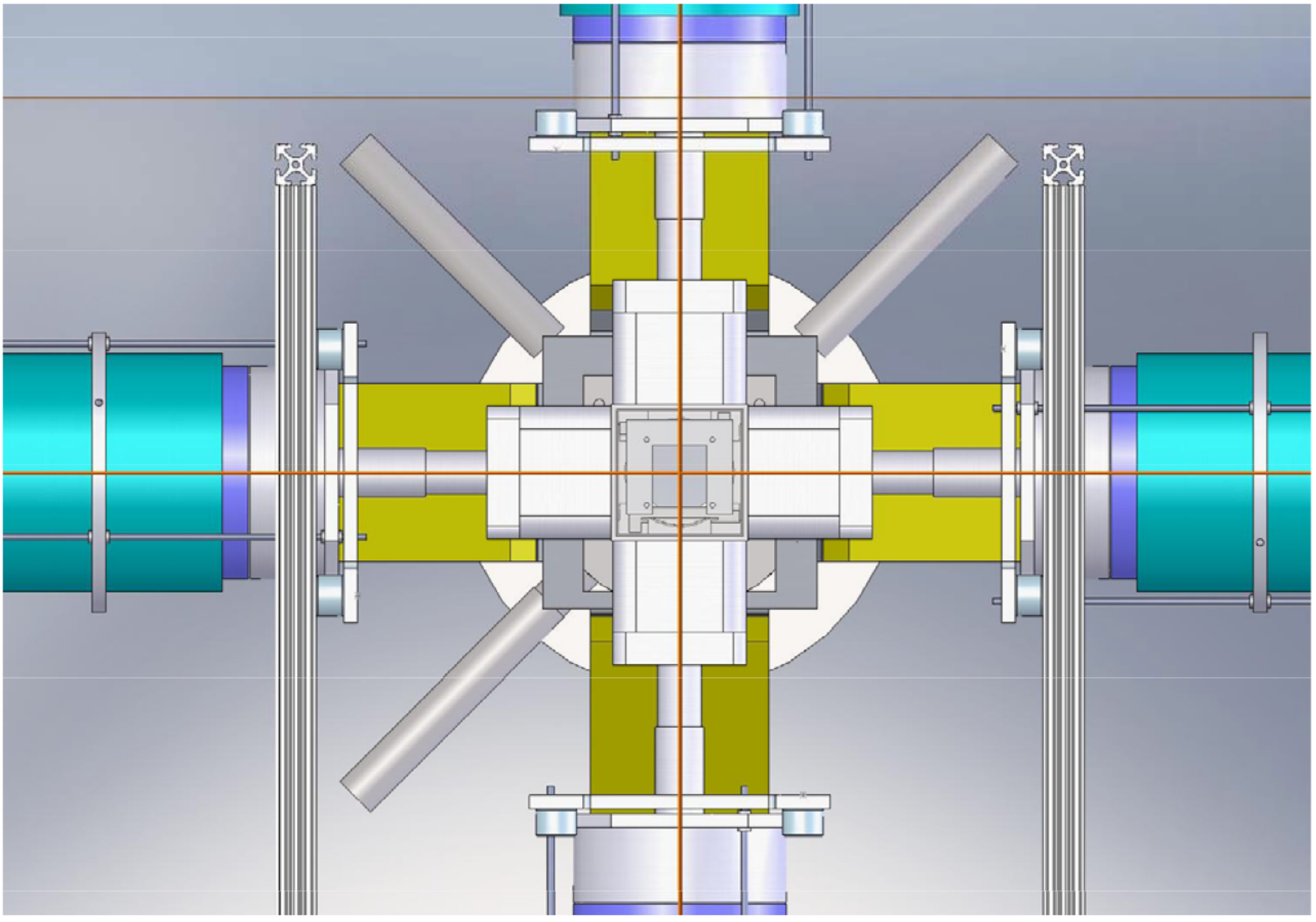}
\caption{\label{fig:xarray} (Color online) Technical diagram of the X-Array of HPGe detectors as described in the text.}
\end{center}
\end{figure}

The X-Array consists of five HPGe clover detectors constructed by Canberra/Eurisys \cite{canberra}. Four identical clovers (160$\%$) consisting of four 60~mm $\times$ 60~mm crystals each are positioned in the vertical plane. The remaining clover (230$\%$), the `Super-Clover', located in the horizontal plane facing the incoming beam, consists of 70~mm $\times$ 70~mm crystals. The detectors are mounted on an adjustable 80/20 aluminum frame with wheels to allow it to be easily moved from one location in the laboratory to another. The position of each detector is adjustable to allow for precise alignment with the experimental set-up. With the flat edges of each detector in the vertical plane in contact with their adjacent neighbors, and the Super-Clover positioned so that it is in contact with the other four detectors, the solid angle coverage is $\sim65\%$ of $4\pi$. The X-array is complete with electronics, high-voltage supply, an automated liquid nitrogen auto-fill system and an emergency bias shutdown system. A technical diagram of the X-Array as used with the DSSD is shown in Fig.~\ref{fig:xarray}. 

\subsection{The plastic scintillator detectors}\label{plastic}

\begin{figure}[h!]
\begin{center}
\includegraphics[width=8.1cm]{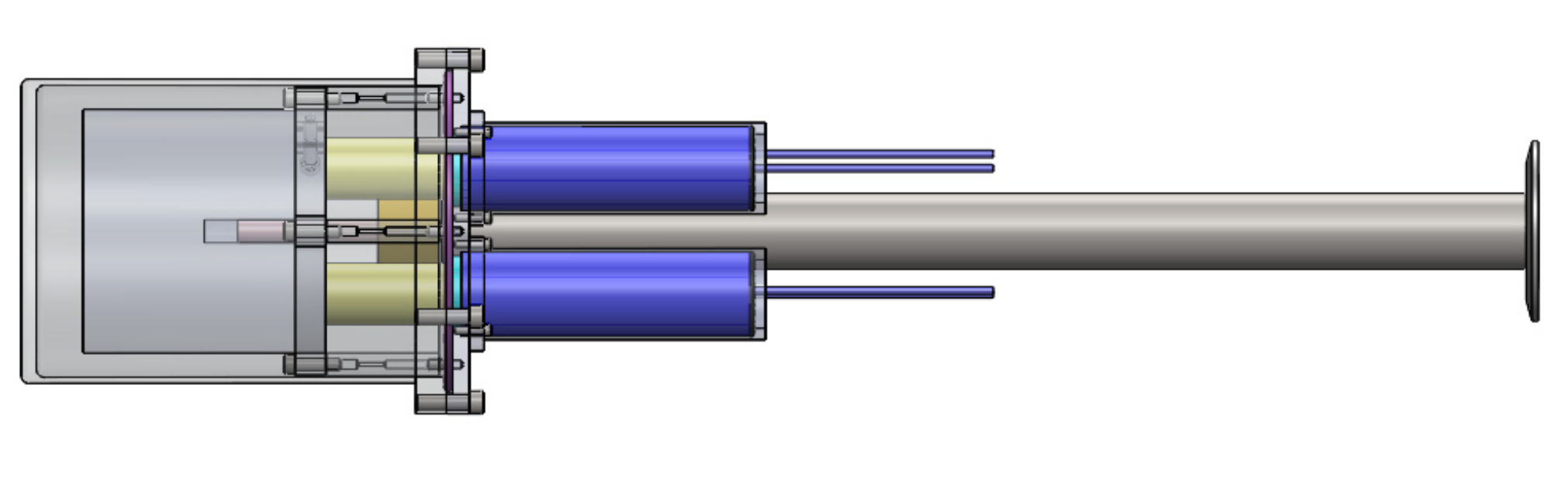}
\includegraphics[width=9cm]{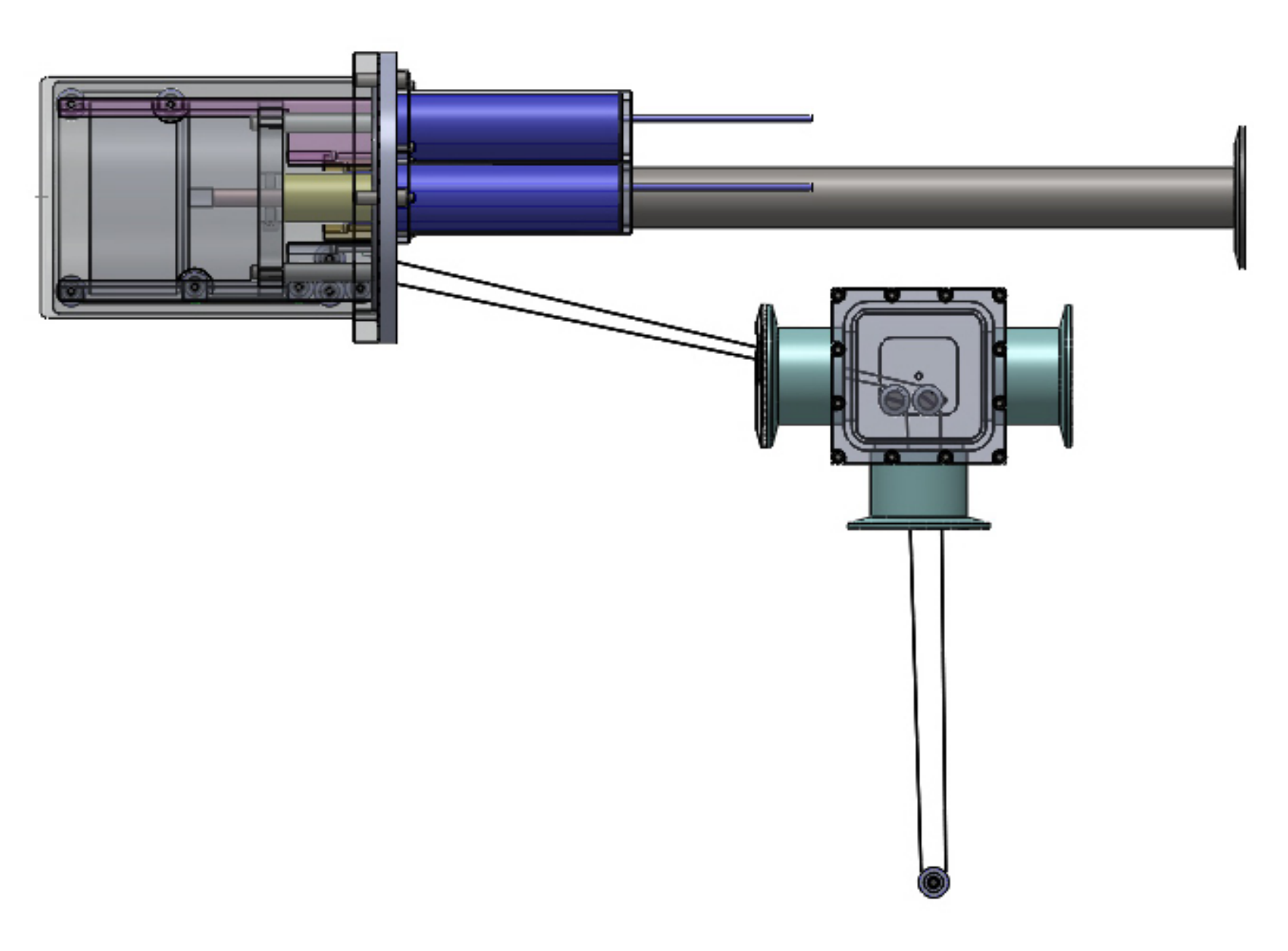}
\includegraphics[width=8.2cm]{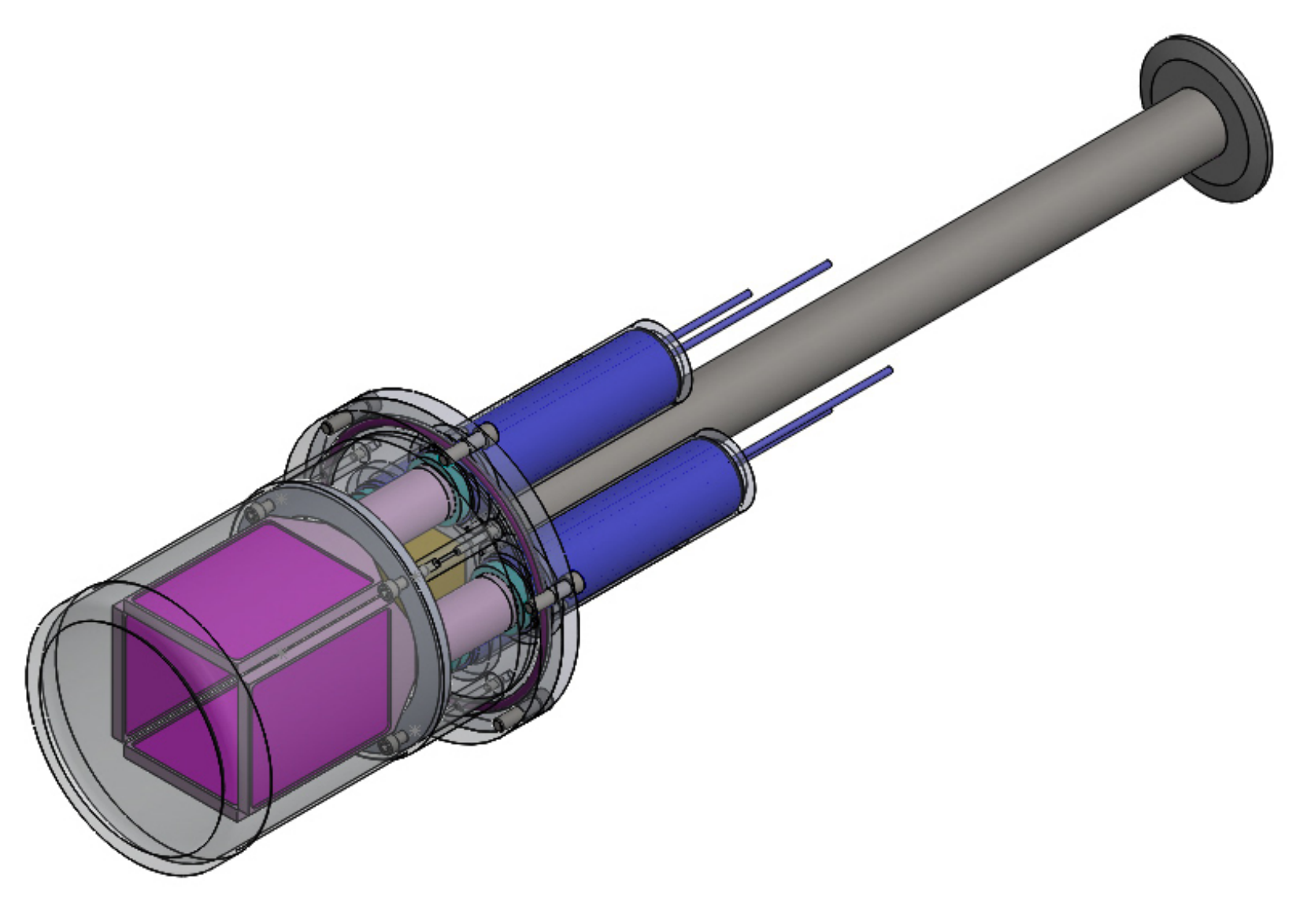}
\caption{\label{fig:plastic} (Color online) Technical drawings of the various $\beta$-particle detectors available: (Top) The Mark-1 chamber with large-volume plastic detector; (Middle) The modified large-volume plastic detector for SATURN; (Bottom) The Mark-1 chamber and paddle system.}
\end{center}
\end{figure}

The scintillator detectors are fabricated from BC-408 plastic (Saint Gobain \cite{saintgobain}). The well detector is a 5-cm diameter cylinder 10 cm in length. The incoming beam enters, via a brass collimator, through a 6-mm diameter well which has been bored in the plastic. Radioactivity is collected on aluminum foil positioned at the bottom of the well. Symmetrically-positioned, 25-mm diameter cylindrical light guides 4~cm long, made from the same material, are coupled to the flat face of the block. The surface of the plastic is coated with TiO reflective paint to improve light collection. Scintillation light is transmitted towards borosilicate optical windows fixed in the back plate of the vacuum chamber. Photomultiplier tube (PMT) modules (Hammamatsu h10492e \cite{hammamatsu}) coupled to the outer surface of the windows are held in place by custom-made housing. Each PMT module contains a high-voltage converter, thus only requiring low-voltage input ($\sim 0.5-1.0~V$). A custom-made, adjustable, low-voltage power supply has been installed to provide power to each individual PMT. The PMT output energy signals are passed via a series of scintillator preamplifiers directly into separate channels in the data acquisition system. A second well detector has been constructed with modifications to allow collection of activity on the moving tape system. The well detector and separate light guides can replaced by the paddles. Each paddle is a 5~cm $\times$ 10~cm $\times$ 6~mm rectangular block which is connected to a cylinder of similar dimensions as the previous light guides. Schematic diagrams of the vacuum chambers and plastic detectors are displayed in Fig.~\ref{fig:plastic}. 

The scintillators are held within an aluminum `top-hat' shaped vacuum chamber. The walls of the chamber are 6~mm thick and designed to maintain mechanical strength under vacuum while minimizing $\gamma$-ray attenuation. Monte Carlo simulations of the detection system performed with MCNP5 version 1.60 \cite{Brown:2010} were used to guide modifications to the original design which improve $\gamma$ efficiency; 3.8 mm of material was removed from the front face of the original chamber and 1.3 mm was removed from a 10-cm wide band corresponding to the location of the vertical clover detectors. The steel back plate of the Mark-1 chamber has four symmetrically-spaced glass windows, as described above, and a 45-cm long, 4-cm diameter beam pipe with NW50 flange for coupling to the CARIBU beamline. The second chamber for SATURN is modified to allow an entry and exit point for the tape in the backplate, with the addition of a series of pulleys surrounding the plastic to create a continuous circuit for the motion of the tape. 


\subsection{A prototype diagnostic tape station for the stub-line}\label{stubline}

A prototype tape-transport station has been designed, constructed and commissioned at the CARIBU low-energy stub-line. The design of this system was based on the original Louisiana State University design \cite{Mlekodaj:1981}, modified to improve speed and efficiency and is equipped with two 30$\%$ HPGe detectors and a plastic scintillator for $\beta$-particle detection. The stub-line tape station is available for beam diagnostics and testing as care was taken to ensure the base pressure of the tape box could be maintained `in vacuum' at pressure $< 10^{-6}$~Torr.

\subsection{The tape-transport system}\label{tape}

\begin{figure}[h!]
\begin{center}
\includegraphics[width=8.5cm]{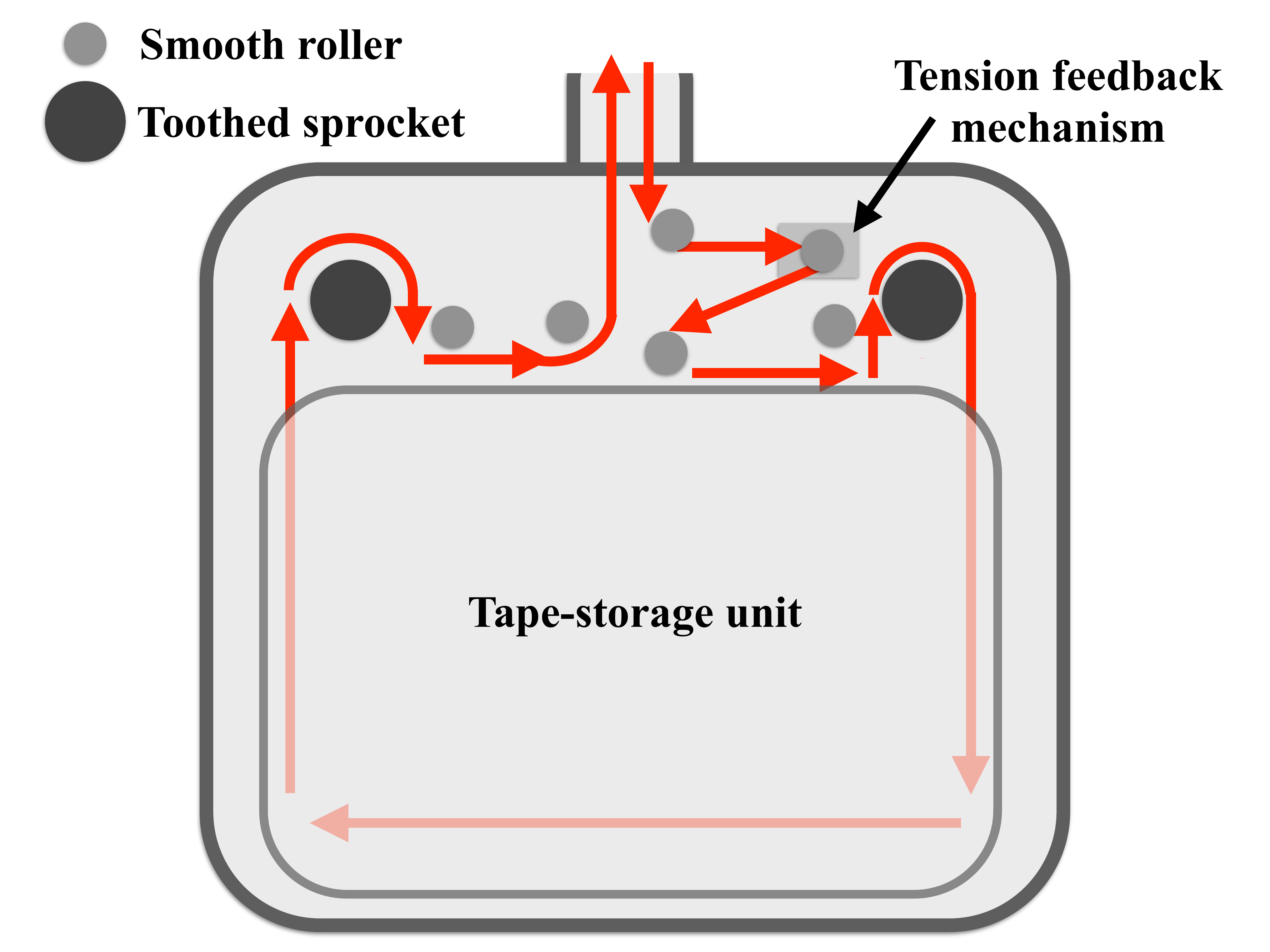}
\caption{\label{fig:drivesystem} (Color online) Schematic diagram of the pulley system that drives the tape and the cassette storage unit. The arrows indicate the direction of motion as the tape passes through the system.}
\end{center}
\end{figure}

The design of SATURN is developed from the aforementioned diagnostic tape station. Two pumping systems are in place: one on the beamline and one on the main tape box, both of which are identical in their components. Vacuum pressure $<$ $\sim 1 \times 10^{-6}$ Torr is achieved in operation via continuous pumping by XDS-10 scroll pumps (Edwards \cite{edwards}) and turbo pumps (Oerlikon Leybold \cite{leybold}). The tape system is based upon a continuous loop of 35-mm aluminized mylar tape. The cassette housing in which the tape is stored is capable of storing  over 100 m of tape. The tape is driven by a OMHT23-400 stepping motor (Omega \cite{omega}) which is mounted outside the vacuum chamber. The drive shaft of the stepping motor is connected to the pulley system via a ferrofluidic rotary feedthrough (Ferrotec USA Corp. \cite{ferrotec}). A schematic diagram of the pulley system that is used to drive the tape is provided in Fig.~\ref{fig:drivesystem}.  

The tape follows a clockwise motion within the vacuum chamber. It exits the storage cassette to the left, passing over a toothed sprocket whilst being guided by adjustable pulleys towards a vertical shaft. At the end of this shaft, it is directed towards the scintillator chamber by a another set of pulleys within a T-shaped vacuum box. When the tape enters the scintillator chamber, it moves below the plastic, vertically along the back face of the chamber, then back towards the centre where it is directed down through a vertical slot in the plastic. At the centre of this slot lies the point at which radiation is deposited on the tape. The tape continues through the entire width of the plastic, where it rounds another pulley and moves back out of the scintillator chamber. From here, it passes over a second pulley within the T-box, back down the vertical shaft and into the main chamber. At this point, the tape is deflected to the right where it passes over a pulley mounted on a movable sliding arm, held in place by a spring. In addition to maintaining tension in the tape throughout the system, this provides an alarm if the tape were to break. The sliding arm is located next to a microswitch that feeds back into the motor control. If tension in the tape is lost, the switch is closed, sending a shut-down signal to the motor controller. The tape is fed back into the cassette via another toothed sprocket at the right-hand side of the chamber. The entire system is held under vacuum during operation and can be isolated from the CARIBU beamline by a gate valve. 

\begin{figure}[h!]
\begin{center}
\includegraphics[width=8.5cm]{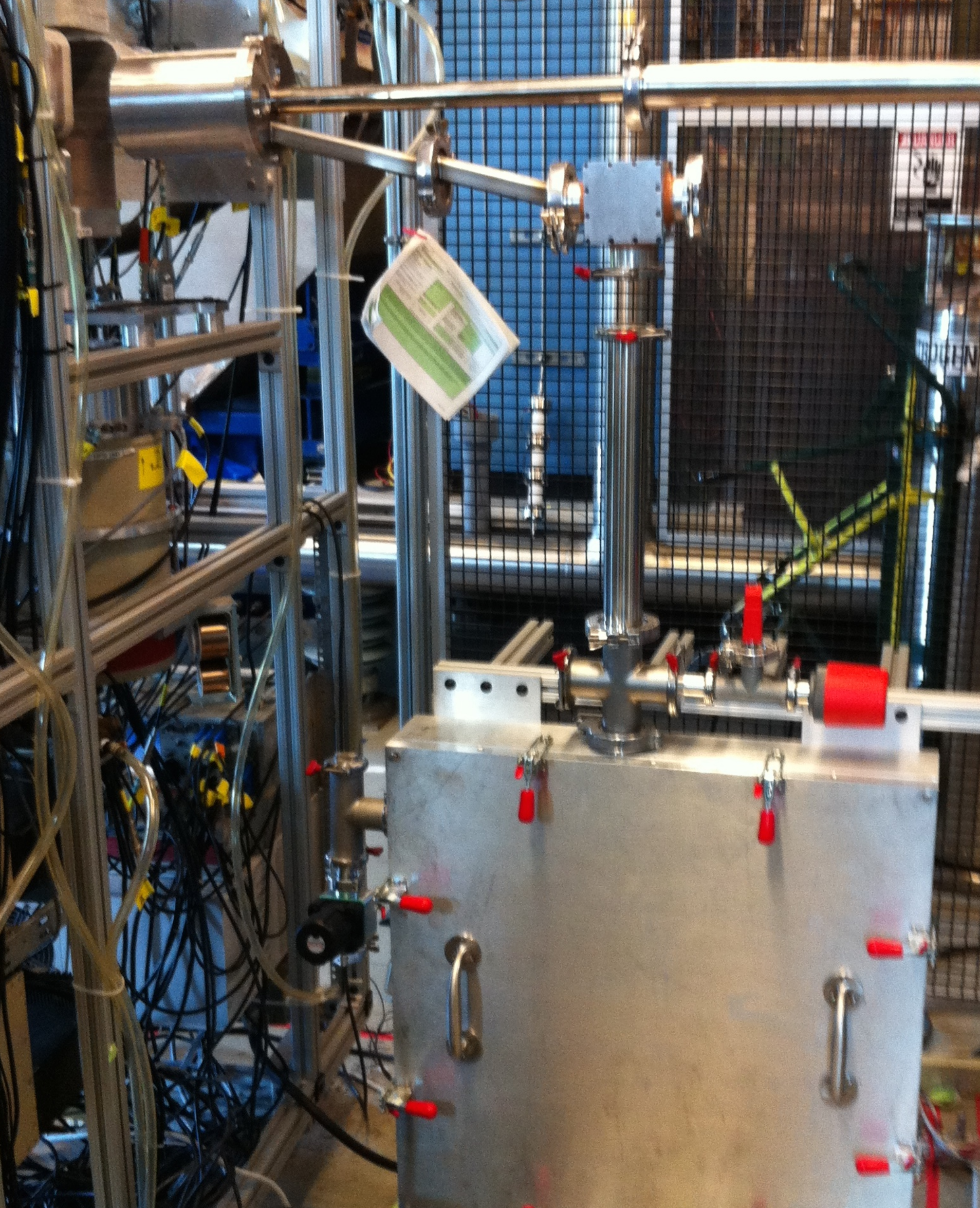}
\caption{\label{fig:tape} (Color online) Photograph of the X-Array and SATURN installed at CARIBU.}
\end{center}
\end{figure}

The stepping motor is operated by a STAC6-Si Programmer (Omega). This is a stand-alone programmable drive in which inputs and outputs are controlled via software installed on a laptop. With this software it is possible to control the distance of tape movement, speed and amount of time taken to move the tape to high precision. The Programmer also allows for control via serial commands sent from a LC880 Programmable Logic Controller (PLC) (LabSmith \cite{labsmith}). A photograph of the full decay station is shown in Fig.~\ref{fig:tape}. 

\subsection{The digital data acquisition system}\label{daq}

The data acquisition system is based upon the recently-implemented Digital Gammasphere system \cite{Anderson:2012}. At the heart of this are four GRETINA (Gamma Ray Energy Tracking In-Beam Nuclear Array) style digitizer modules \cite{Anderson:2009} which digitize the Ge preamp waveforms at a 100-MS/s rate. The digitizer boards provide 10 channels of 14-bit flash ADC each, which are read out by a single Xilinx XC3S5000 FPGA. Firmware for the digitizer modules, specific to the needs of the array, have been developed. The clock distribution design ensures the synchronization of ADC clocks of each individual channel. Incoming data enters a delay pipeline consisting of multiple buffers within the FPGA. A four-stage Gaussian filter is applied to the leading-edge discriminator output, which is in turn compared to a user-defined threshold. With the firing of the discriminator, the continuous running sums across the `pre-rise' and `post-rise' buffers are sampled. The difference of these corresponds to the height of the input pulse. New firmware for the trigger modules has also been implemented, with the digitizer modules reporting the condition of all 10 channel discriminators to the trigger at 20-ns intervals. The singles trigger for data acquisition permits any of the 20 Ge crystals or a signals from the scintillator plastic to initiate a readout. Currently, energy and time are written for each Ge channel and full traces for the plastic, tape and buncher channels. Data is written to disk at a rate of $\sim 1.0 - 1.5$~MB/second. Each of the 10-channel discriminators is buffered, so there is very little system dead time involved - one channel can be reading out while the other 19 stay active.


\section{System performance}\label{performance}

The $\beta$ decays of $^{142, 144, 146}$Cs $\rightarrow$ $^{142, 144, 146}$Ba were utilized to test the performance of the new decay station. Data were collected with both the Mark-I $\beta$ chamber and SATURN in several commissioning runs. A variety of efficiency measurements were performed using standard sources, in addition to collection of experimental data including $\gamma$ and $\beta\gamma$ spectra, $\gamma\gamma$ and $\beta\gamma\gamma$ coincidence matrices and decay curves. 


\subsection{X-Array `direct-hit' efficiency and resolution}\label{xefficiency}

\begin{figure}[h!]
\begin{center}
\includegraphics[width=8.2 cm]{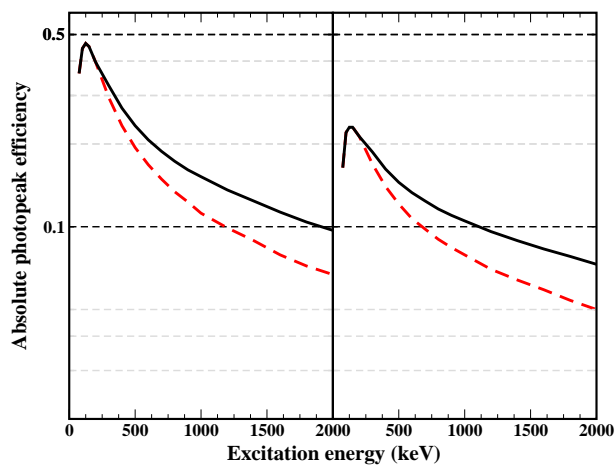}
\caption{\label{fig:simulations} (Color online) Results of MCNP5 simulations for the photopeak efficiency of the X-Array. Left: X-Array only; Right: X-Array and large-volume plastic scintillator detector. The dashed lines are for `1-hit' events. The solid lines include multiple-hit events reconstructed from Add-Back.}
\end{center}
\end{figure}

The efficiency performance of the X-Array for detecting $\gamma$ rays up to 2~MeV in excitation has been simulated using MCNP5. In the simulations, the exact dimensions of the crystals and the housing material have been taken into account. The simulated photopeak efficiency of the X-Array operating alone and with the plastic well detector is shown in Fig.~\ref{fig:simulations}. 

\begin{figure}[t!]
\begin{center}
\includegraphics[width=8.2cm]{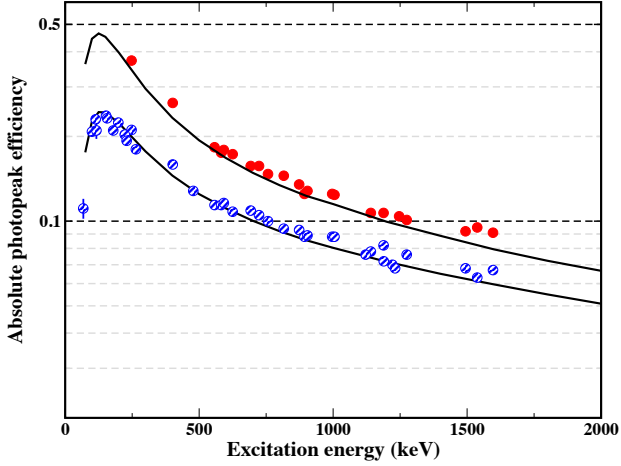}
\caption{\label{fig:efficiency} (Color online) Absolute photopeak efficiencies measured for the X-Array operating alone and with the plastic well detector, as described in the text. Data points are the measured efficiencies (fully shaded for X-Array alone, partially shaded for X-Array and $\beta$-particle detector), and the solid lines are MCNP5 simulations. $^{154}$Eu data were collected for both configurations. Additional $^{182}$Ta data were collected for the full $\beta$-detector system.}
\end{center}
\end{figure}

Absolutely-calibrated $^{154}$Eu ($\sim$5~$\mu$Ci) and $^{182}$Ta ($\sim$~30 nCi) sources were used to measure the $\gamma$-ray detection efficiency of the X-Array. This was performed by building up the following layers in the detection system: X-Array only; X-Array plus the Al vacuum chamber; X-Array plus the modified large-volume plastic scintillator; and the X-Array plus the full scintillator chamber and plastic used with SATURN. This was achieved by inserting the sources into the centre of the plastic. The results of these efficiency measurements and comparison with the simulations are provided in Fig.~\ref{fig:efficiency}. The absolute efficiency of the X-Array was found to be $\sim50\%$ at 122~keV and $\sim10\%$ at 1332~keV. The Al vacuum chamber attenuates $\gamma$ rays by $\sim10\%$. The combined system has a photopeak efficiency of $\sim25\%$ at 122~keV and $\sim 5\%$ at 1332~keV. 

The energy resolution, corresponding to the full-width half-maximum (FWHM) of individual crystals, clovers and the full X-Array has been examined using $^{146}$Cs decay data collected in commissioning. The performance for the X-Array at 1332 keV is 3.2-keV FWHM. This is considerably worse than the intrinsic array response ($\sim$ 2.3~keV) and points to the need for further optimization of the digital algorithms. 

\subsection{X-Array cross scattering and summing}\label{xaddback}

\begin{figure}[h!]
\begin{center}
\includegraphics[width=8.2cm]{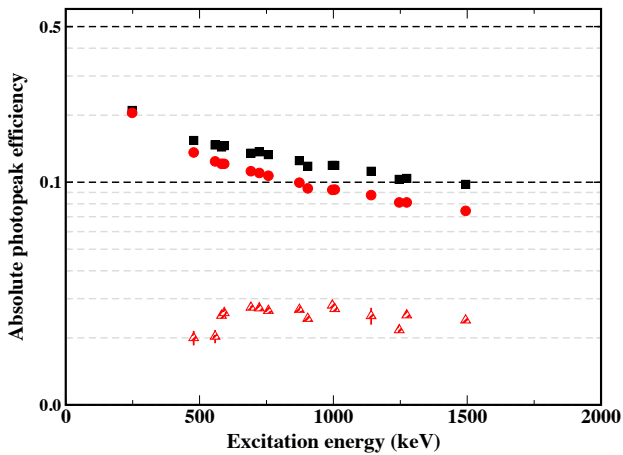}
\caption{\label{fig:addback} (Color online) Absolute photopeak efficiencies for direct-hit (squares), two-hit (triangles) and add-back (circles) photopeak efficiencies measured for the X-Array and $\beta$-chamber combination.}
\end{center}
\end{figure}

As discussed above, with large clover detectors incident $\gamma$ rays have a probability of Compton scattering between crystals within each clover and depositing their energy in more than one crystal, and between individual clovers. The `add-back' technique, which is discussed in greater detail in Ref.~\cite{Duchene:1999}, may be used in off-line data sorting to increase photopeak efficiencies by summing partial energies of $\gamma$ rays that deposit their full energy between adjacent crystals. Direct-hit, two-hit and full add-back detection efficiencies measured for the X-Array and $\beta$-chamber combination are presented in Fig.~\ref{fig:addback}. 

For low $\gamma$-ray energies, Compton scattering effects are expected to be small, hence the direct hit efficiency is similar to the add-back efficiency below 200-keV. Above 200~keV, the probability of scattering between crystals increases and a greater number of double hits are reconstructed. The add-back factor, $F(E_{\gamma})$, obtained from the reconstructed multiple-hit events varies linearly with the logarithm of $\gamma$-ray energy and has the form:

\begin{center}
\begin{equation}
F(E_{\gamma}) = 0.18ln(E_{\gamma}) - 0.93,
\end{equation}
\end{center}

\noindent
and has value of 0.37 at 1332~keV.


\subsection{Plastic scintillator efficiency and tape reliability }\label{befficiency}

The detection efficiency of the large-volume plastic scintillator detector has been estimated using $^{146}$Cs data. The number of counts in photopeaks in $\gamma$ `singles' spectra were compared to the number of counts in the corresponding $\beta\gamma$ spectra photopeaks. From this approach, the efficiency of the detector is estimated to be 60-70$\%$. 

Mobility and reliability of the tape motion was tested over a period of 24 hours with repeated 6-second cycles. In each cycle, the tape was moved by $\sim$1~m (the distance from the implantation point within the scintillator detector to the point just inside the main tape storage unit) in a period of $< $1 second. During this time, the tape motion showed no signs of deteriorating. After insertion through the system, loose ends of the tape are welded together using an ultrasonic splicer. This vastly improves the strength of the tape over other more conventional means of connecting the loose ends. Electronics and experiment control mechanisms have been fully implemented during testing. In commissioning, both the SATURN logic controller and CARIBU buncher were demonstrated to be capable of providing a suitable master clock for controlling beam cycles. Both systems are available depending on the requirements of the experiment and half-life of the nucleus of interest.


\section{Commissioning}\label{commissioning}

\subsection{Mark-I scintillator detector}\label{mark1}
\begin{figure}[h!]
\begin{center}
\includegraphics[width=8.2cm]{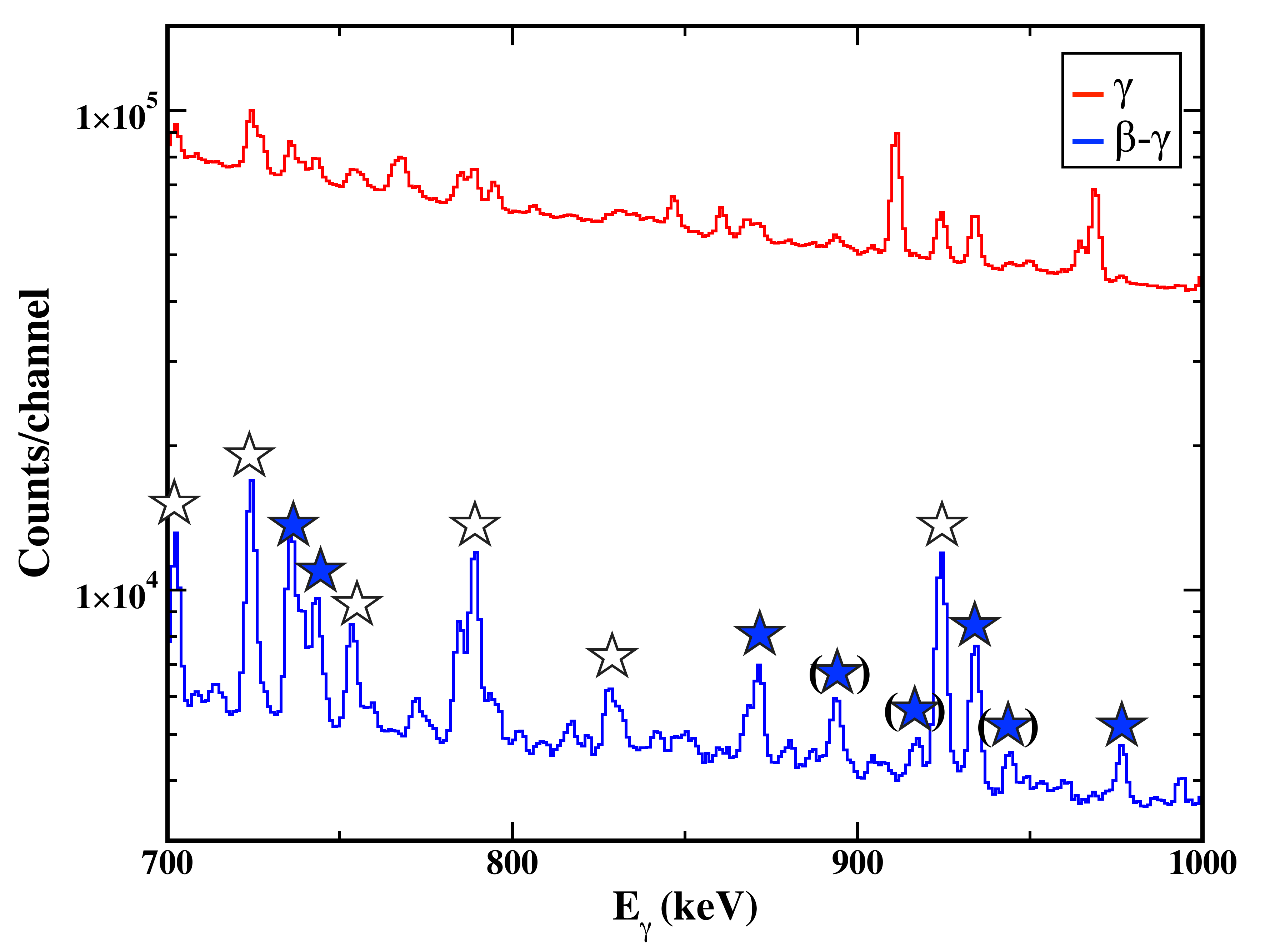}
\caption{\label{fig:a146Ba_spectrum} (Color online) 700 keV - 1000 keV region of the $\gamma$ (upper) and $\beta\gamma$ (lower) singles spectra recorded with a $^{146}$Cs beam using the Mark-I chamber. $\gamma$-ray transitions in $^{146}$Ba are marked with filled stars. Transitions from other members of the $A~=~146$ decay chain are marked with unfilled stars. Transitions that are not included in the adopted level scheme are indicated by parentheses.}
\end{center}
\end{figure}

The Mark-I chamber, featuring the large-volume plastic scintillator detector, was used to collect raw $\gamma$-ray singles events, $\beta\gamma$ and $\beta\gamma\gamma$ coincidence data using $^{142,146}$Cs ions from the CARIBU source. Examples of the  $\gamma$ singles and $\beta$-gated $\gamma$ singles spectra that were recorded for $^{146}$Cs decay are provided in Fig.~\ref{fig:a146Ba_spectrum}. Timing conditions for coincidence events are applied in off-line data sorting. The advantage of possessing the ability to gate upon coincident $\beta$ particles is evident. Spectral background is suppressed by an order of magnitude, resulting in enhancement of the $\gamma$ rays of interest. In these measurements, the Cs beam was deposited on an aluminum foil at the end of the brass collimator at the geometric centre of the plastic scintillator. Despite the foil being replaced periodically, long-lived members of the $A=146$ decay chain accumulated over time and observed in the data. 

\begin{figure}[h!]
\begin{center}
\includegraphics[width=8.2cm]{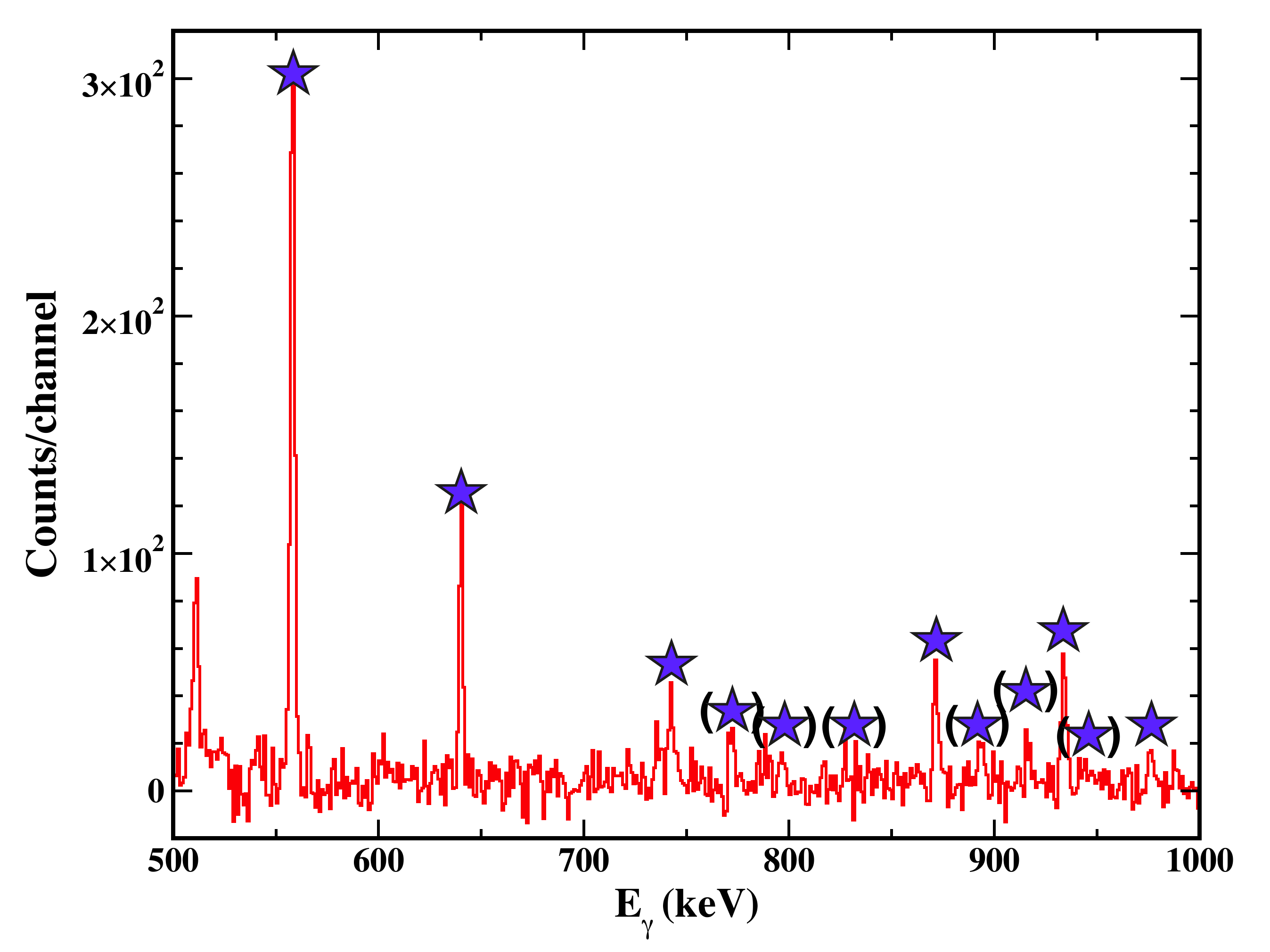}
\caption{\label{fig:a146Csggx181} (Color online) 500 keV - 1000 keV region of the background-subtracted projection of the $\beta\gamma\gamma$ matrix, gated on the 181-keV 2$^+_1\rightarrow0^+_{{\rm g.s.}}$ transition in $^{146}$Ba. Transitions that are not included in the adopted level scheme are indicated by parentheses.}
\end{center}
\end{figure}

$\gamma\gamma$ and $\beta$$\gamma\gamma$ coincidence events were also recorded. Fig.~\ref{fig:a146Csggx181} shows a background-subtracted projection of the $\beta\gamma\gamma$ matrix that has been gated on the 181-keV 2$^+_1\rightarrow0^+_{{\rm g.s.}}$ transition in $^{146}$Ba \cite{Peker:1997}. Transitions that would be expected from the existing level scheme are indicated with blue stars. A number of weak transitions identified in the data to be in coincidence with the 181-keV transition, and are not included in the adopted level scheme, are labelled with parentheses. Such coincidence data are used in identifying the origin of unknown $\gamma$ rays. Transitions in all members of the $^{146}$Cs $\rightarrow$ $^{146}$Ba  $\rightarrow$$^{146}$La  $\rightarrow$$^{146}$Ce  $\rightarrow$$^{146}$Pr  $\rightarrow$ $^{146}$Nd decay sequence have also been identified. A number of the most intense $\gamma$ rays from $\beta$-delayed neutron emission of $^{146}$Ba $\rightarrow$ $^{145}$Ba have also been observed. \\

\subsection{SATURN}\label{mark2}

Commissioning of SATURN was conducted using beams of $^{144,146}$Cs from the CARIBU source. The Mark-I chamber has four PMT modules, whereas SATURN has three. We did not observe any dramatic reduction in light-collection efficiency from the scintillator, which was modified from the first design to allow for motion of the tape. In addition to the raw $\gamma$-ray singles events, $\beta\gamma$ and $\beta\gamma\gamma$ coincidence data provided by the Mark-I chamber, timing information relative to the clock set by the tape control or buncher is included. 

\begin{figure}[h!]
\begin{center}
\includegraphics[width=8.2cm]{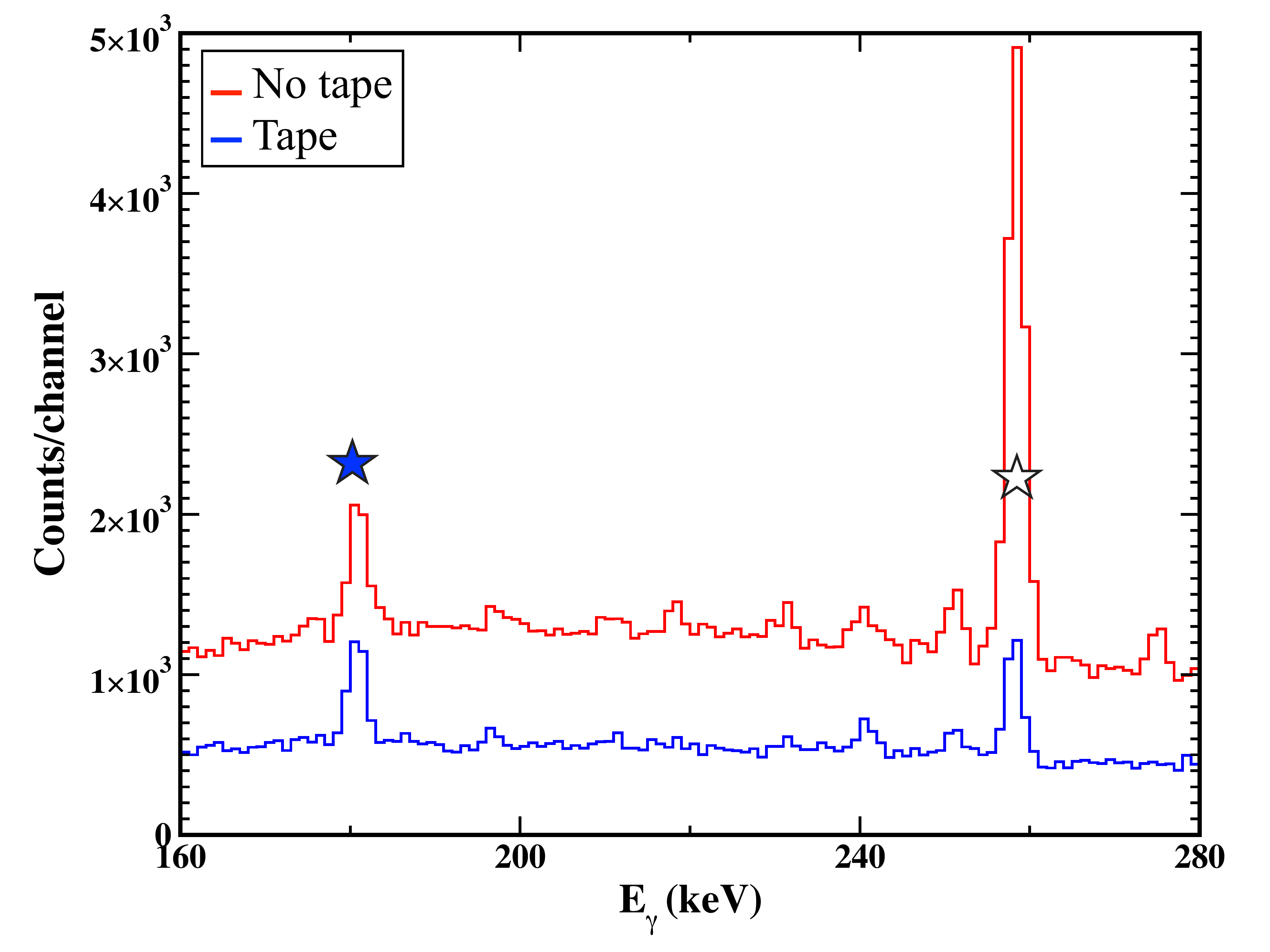}
\caption{\label{fig:s146Cs_TNT} (Color online) Comparison of data collected with and without the tape transport system. The filled star indicates the 181-keV 2$^+_1\rightarrow0^+_{{\rm g.s.}}$ transition in $^{146}$Ba. The unfilled star indicates the 258-keV 2$^+_1\rightarrow0^+_{{\rm g.s.}}$ transition in $^{146}$Ce.}
\end{center}
\end{figure}

The main role of the tape is to transport long-lived radiation away from the detectors to avoid build up of activity over time. Data collected with a $^{146}$Cs beam, with and without the tape motion, is compared in Fig.~\ref{fig:s146Cs_TNT}. The blue star indicates the 181-keV 2$^+_1\rightarrow0^+_{{\rm g.s.}}$ transition in $^{146}$Ba. The yellow star highlights the 258-keV 2$^+_1\rightarrow0^+_{{\rm g.s.}}$ transition in the long-lived contaminant $^{146}$Ce. By moving the tape, the measured yield of the isobaric contaminant and background is significantly reduced with respect to the yield of $\gamma$ ray from  $^{146}$Ba.

\vspace{-0.01cm}


\section{Half-life measurements}\label{data}

The timing information provided from the beam sweep and tape system allows the measurement of decay half-lives. Microsecond timestamps of $\gamma$ and $\beta$ events are sorted relative to the master clock. The timing cycle is determined by the experiment. Each cycle commences with a short period of background collection. A pulse sent to the kicker signals the start of the collection period. The beam is deflected away from the experimental beamline during the decay period. The cycle is completed with a tape move and final background collection. 

\subsection{Stub-line tape station}

The stub-line tape station was commissioned with $^{142}$Cs ions from the 30-mCi CARIBU source. The time cycle used in measuring the decay half-life is provided in Table~\ref{table:cycle1}. 

\begin{table}[th]
\begin{center}
\caption{Time cycle used for $^{142}$Cs (T$_{1/2}$ = 1.684(14)~s \cite{Johnson:2011}).}
\begin{tabular}{cc}
\hline
Time (seconds) & Action \\
\hline
0 - 3  & Background collection \\
3 - 9  & Beam on (growth) \\
9 - 21  & Beam off (decay) \\
21 - 22  & Tape move \\
22 - 25  & Background collection \\
\hline
\end{tabular}
\label{table:cycle1}
\end{center}
\end{table}

\begin{figure}[h!]
\begin{center}
\includegraphics[width=8.0cm]{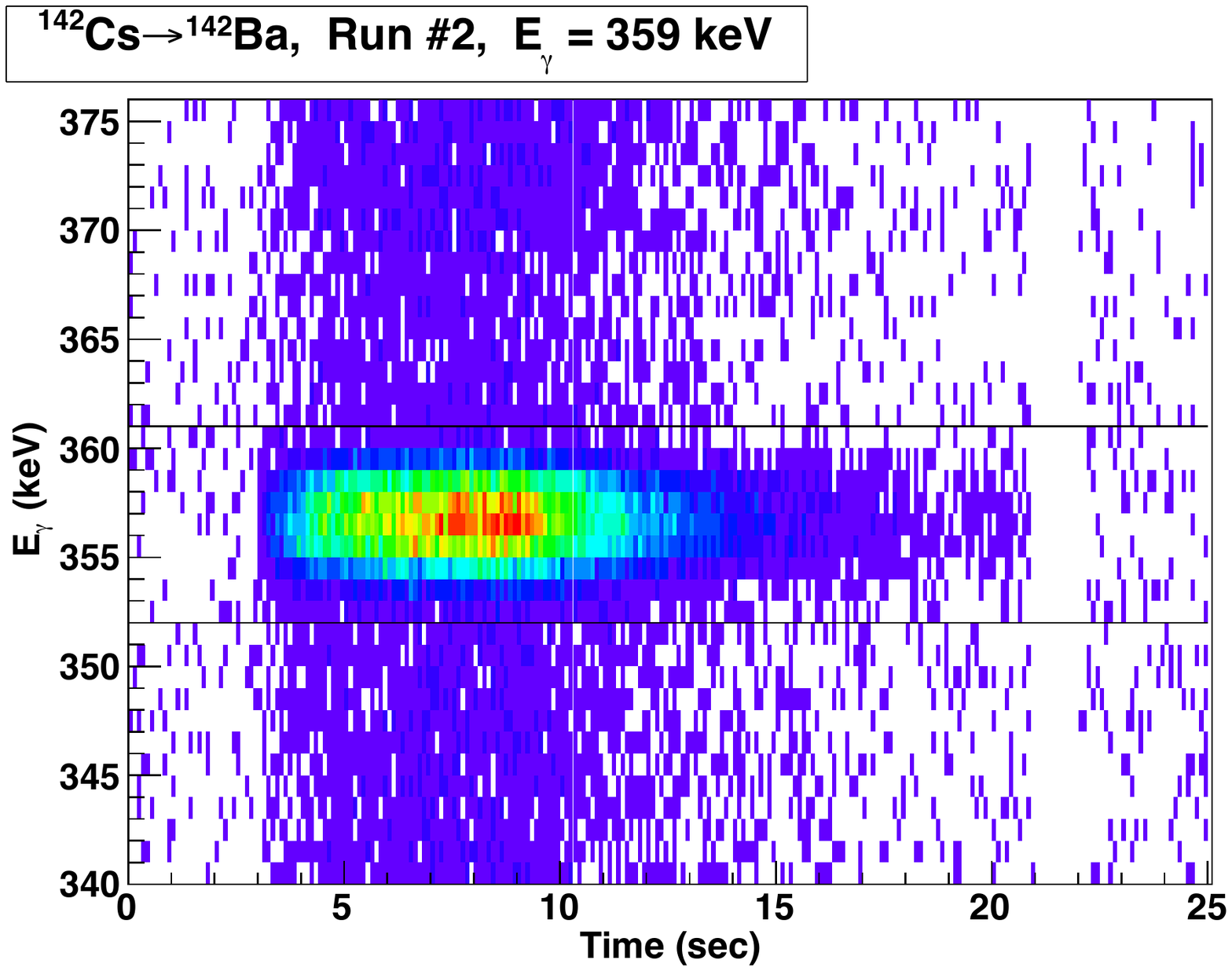}
\includegraphics[width=8.0cm]{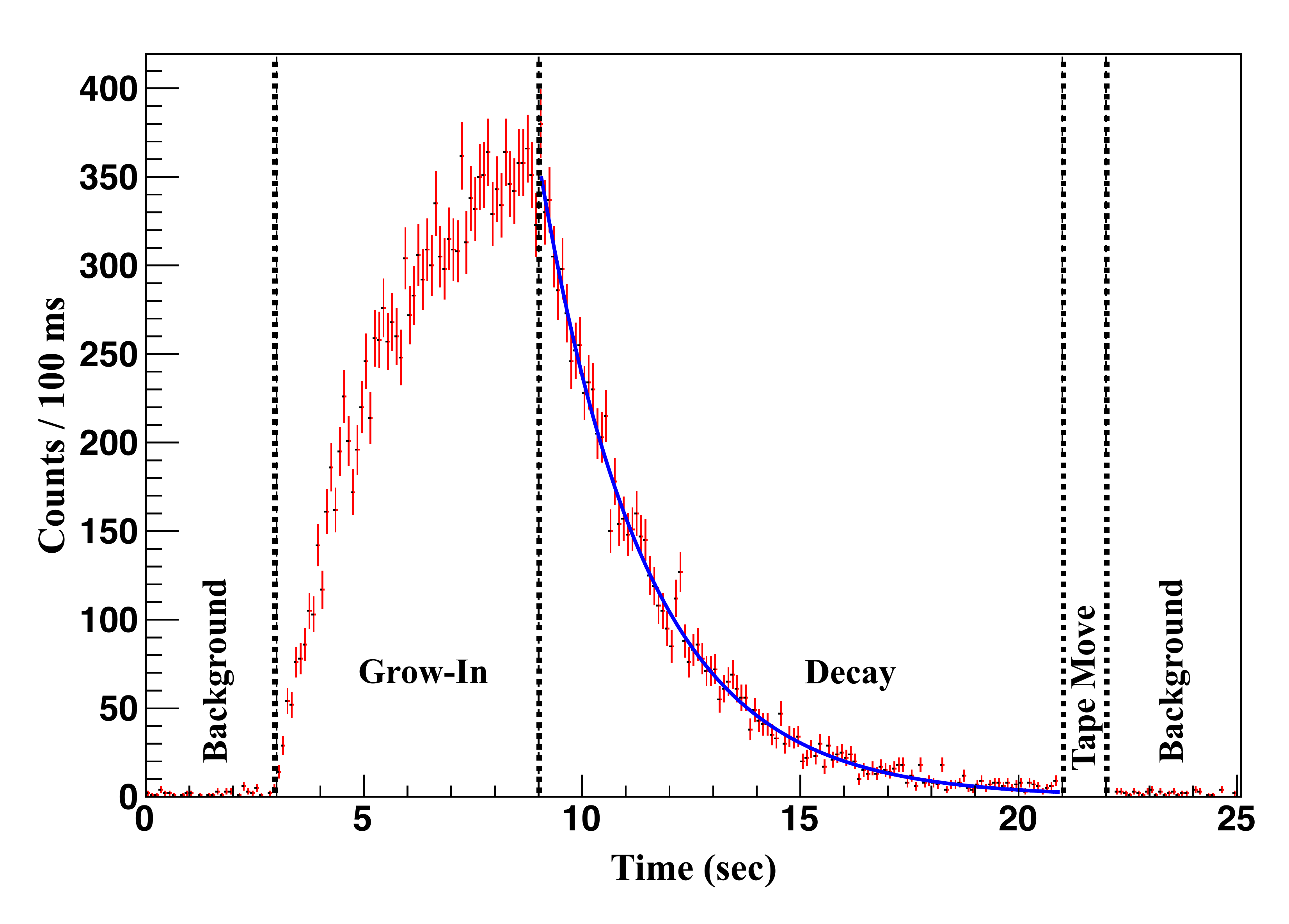}
\caption{\label{fig:s142Cs_RGTg_pb} (Color online) (Top) $\gamma$-ray Energy versus Time spectrum for $^{142}$Cs. (Bottom) Decay curve gated on 359-keV $\gamma$ ray in $^{142}$Ba.}
\end{center}
\end{figure}

A $\gamma$-ray Energy versus Time spectrum is shown in Fig.~\ref{fig:s142Cs_RGTg_pb}. This has been $\beta$-gated and focuses on the 359-keV 2$^+_1\rightarrow0^+_{{\rm g.s.}}$ transition in $^{142}$Ba. By projecting this onto the timing axis a decay curve is obtained. Using such a method it is possible to extract a half-life for the $\beta$ decay of $^{142}$Cs from any $\gamma$ ray that is identified in the data, with minimal background contribution. The $^{142}$Cs $\rightarrow$ $^{142}$Ba half-life was measured to be 1.68(2)~s. This shows excellent agreement with the adopted values of 1.684(14)~s \cite{Johnson:2011}, thus demonstrating the ability of the system to accurately measure decay half-lives. Decay curves obtained from gating on different $^{142}$Ba $\gamma$ rays yielded consistent results.

\subsection{SATURN}\label{saturn}

\begin{figure}[h!]
\begin{center}
\includegraphics[width=8.0cm]{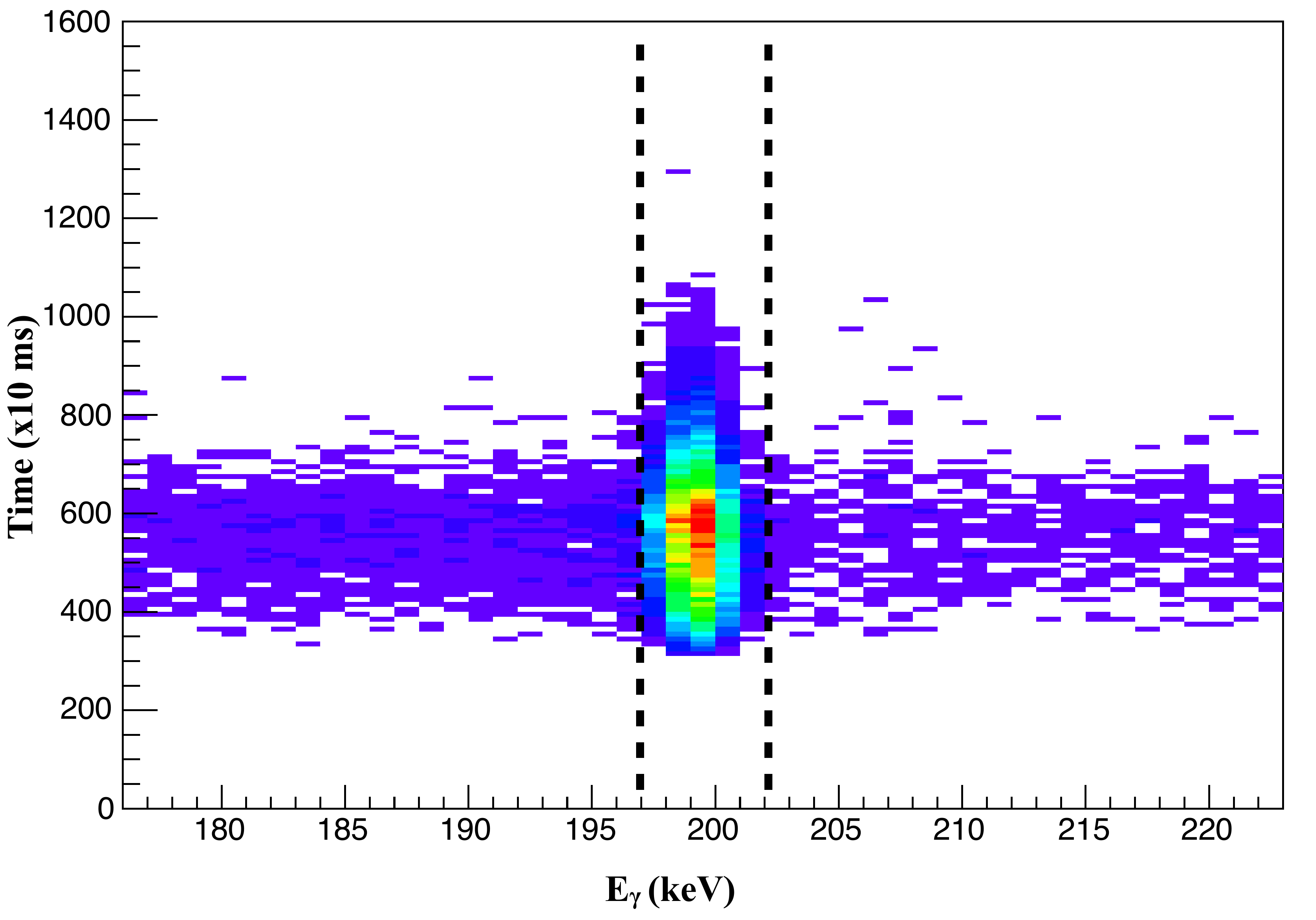}
\includegraphics[width=8.0cm]{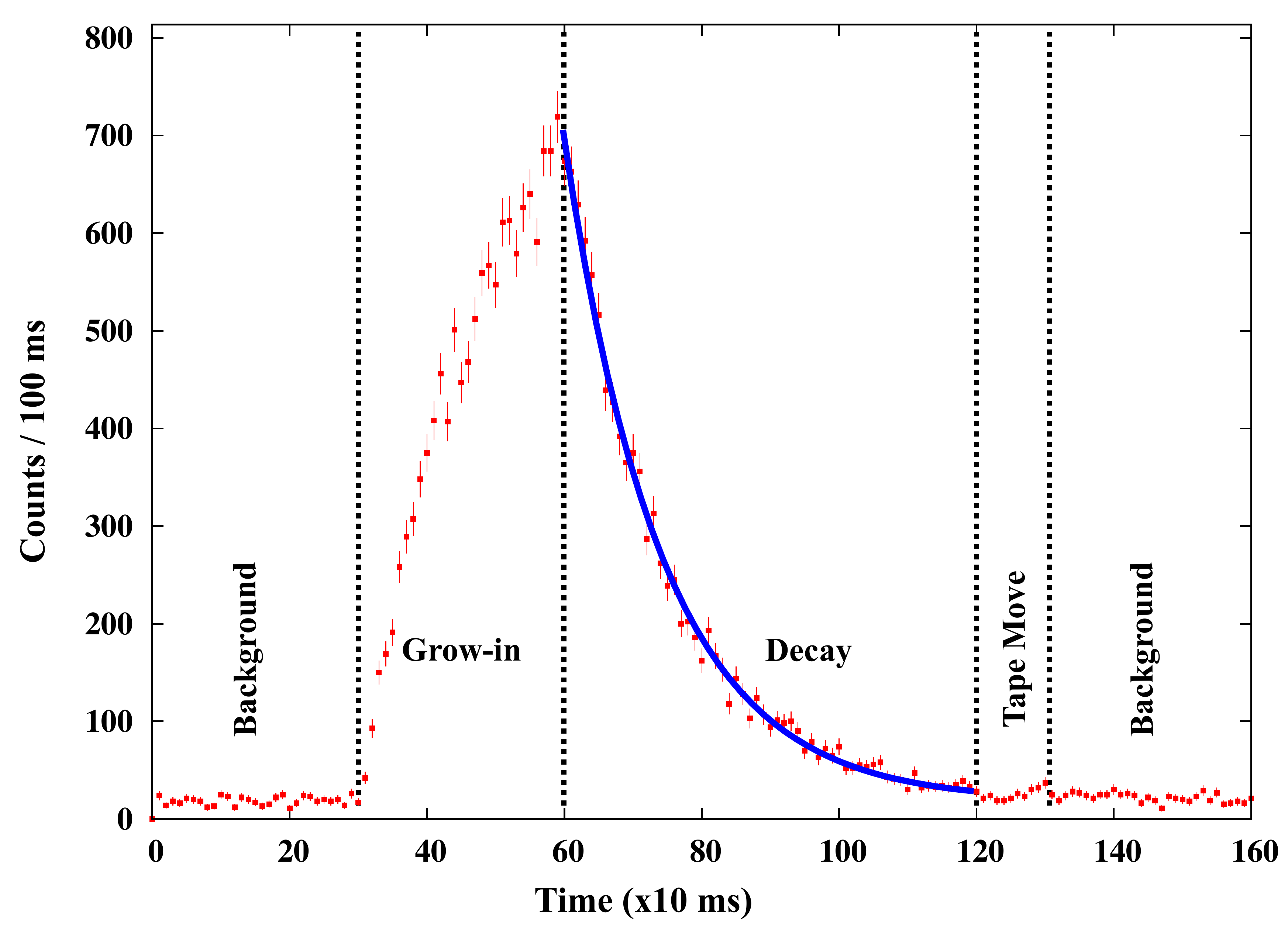}
\caption{\label{fig:s144Cs_RGTg_199} (Color online) (Top) Time versus $\gamma$-ray Energy spectrum for $^{144}$Cs. (Bottom) Decay curve gated on 199-keV $\gamma$ ray in $^{144}$Ba.}
\end{center}
\end{figure}

The time cycle adopted to measure the $^{144}$Cs $\rightarrow$ $^{144}$Ba half-life with SATURN is provided in Table~\ref{table:cycle2}. A Time versus $\gamma$-ray Energy spectrum is shown in Fig.~\ref{fig:s144Cs_RGTg_199}. This has been $\beta$-gated and reduced to focus on the 199-keV 2$^+_1\rightarrow0^+_{{\rm g.s.}}$ transition in $^{144}$Ba. The decay half-life was measured to be 1.00(4)~s, which is within errors of the adopted value of 0.994(4)~s \cite{Sonzogni:2001}.\\

\begin{table}[h!]
\begin{center}
\caption{Time cycle used for $^{144}$Cs (T$_{1/2}$ = 0.994(4)~s \cite{Sonzogni:2001}).}
\begin{tabular}{cc}
\hline
Time (seconds) & Action \\
\hline
0 - 3  & Background collection \\
3 - 6  & Beam on (growth) \\
6 - 12  & Beam off (decay) \\
12 - 13  & Tape move \\
13 - 16  & Background collection \\
\hline
\end{tabular}
\label{table:cycle2}
\end{center}
\end{table}

\section{Summary}\label{summary}

A new decay station capable of performing $\beta$-decay spectroscopy with low-energy CARIBU beams has been commissioned. The experimental set-up is comprised of the X-Array of HPGe clover detectors and the SATURN $\beta$-particle detector and tape-transport station. The full array has been characterized and tested by measuring the $\beta$ decay of various Cs isotopes. The array is ready to begin an extensive experimental campaign that will explore a variety of aspects of nuclear structure, astrophysics and applied applications. The modularity and versatility of the new set-up will also allow for further experiments with the addition of dedicated neutron detectors in the future. 

\section{Acknowledgements}\label{acknowledgements}

The authors wish to express their thanks to E. Zganjar of Louisiana State University for education on the design of tape-transport systems, the Engineering Department at Argonne National Laboratory and the Submillimeter-Wave Technology Laboratory, University of Massachusetts Lowell. This work was supported by the US Department of Energy, Office of Nuclear Physics, under Contract Nos. DE-AC02-06CH11357 and DE-FG02-94ER40848. \\

\end{document}